\documentclass[%
superscriptaddress,
nofootinbib,
amsmath,amssymb,
aps,
pra,
twocolumn,
]{revtex4-2}

\usepackage{amsmath}
\usepackage{amssymb}
\usepackage{amsfonts}
\usepackage{bm} 
\usepackage{bbm}
\usepackage{braket}
\usepackage{color}
\usepackage{comment}
\usepackage{dcolumn} 
\usepackage{dsfont}
\usepackage{enumerate}
\usepackage{epsfig}
\usepackage{esint}
\usepackage[T1]{fontenc}
\usepackage{framed}
\usepackage{gensymb}
\usepackage{graphicx} 
\usepackage[colorlinks,linkcolor=blue,citecolor=blue,urlcolor=blue,hyperindex,driverfallback=dvipdfm]{hyperref}
\usepackage{indentfirst}
\usepackage{lmodern}
\usepackage{mathrsfs}
\usepackage{mathtools}
\usepackage{multirow}
\usepackage{orcidlink}
\usepackage{psfrag}
\usepackage{pst-all}
\usepackage{soul}
\usepackage{xcolor}
\usepackage{xspace}

\newcommand{\ccpar}[1] {\mathopen{}\left(#1\right)\mathclose{}}
\newcommand{\sqpar}[1] {\mathopen{}\left[#1\right]\mathclose{}}
\newcommand{\clpar}[1] {\mathopen{}\left\{#1\right\}\mathclose{}}

\def\ii{{\rm i}}  \def\ee{{\rm e}}
  \def\kB{{k_{\rm B}}}
  \def\Imm{{\rm Im}}

\newcommand{\pd}[2] {\mathopen{}\frac{\partial#1}{\partial#2}\mathclose{}}

\def\rb{{\bf r}}      \def\vb{{\bf v}}

\def\kb{{\bf k}}    
\def\qb{{\bf q}}      
\def\Eb{{\bf E}}        
\def\Jb{{\bf J}}    
    
\def\vF{v_{\rm F}}  \def\kF{{k_{\rm F}}}  \def\EF{{E_{\rm F}}}  
  \def\kkb{\hat{\bf k}}  \def\fk{f_{\bf k}}  
  
\def\ww{\omega}      

\def\Am{\mathcal{A}}        
  \def\Gm{\mathcal{G}}    
    
\usepackage{times}
\usepackage{xcolor}

\begin{document}

\title{Nonreciprocal plasmonic response of drift-biased two-dimensional metals}

\author{Gonzalo \'Alvarez-P\'erez\,\orcidlink{0000-0002-4633-1898}}
\email[Gonzalo \'Alvarez-P\'erez: ]{gonzalo.alvarezperez@iit.it}
\affiliation{Istituto Italiano di Tecnologia, Center for Biomolecular Nanotechnologies, Via Barsanti 14, 73010 Arnesano, Italy}

\author{Joel~D.~Cox\,\orcidlink{0000-0002-5954-6038}}
\email[Joel~D.~Cox: ]{cox@mci.sdu.dk}
\affiliation{POLIMA---Center for Polariton-driven Light--Matter Interactions, University of Southern Denmark, Campusvej 55, DK-5230 Odense M, Denmark}
\affiliation{Danish Institute for Advanced Study, University of Southern Denmark, Campusvej 55, DK-5230 Odense M, Denmark}

\begin{abstract}
We develop a nonlocal electrodynamic framework for drift-biased two-dimensional electron gases (2DEGs) with Dirac (linear) and parabolic (quadratic) dispersions, deriving closed-form drift-dependent conductivity tensors from the Boltzmann transport equation. We obtain the plasmon dispersion and near-field emission of a point dipole, revealing nonreciprocal propagation in both systems. At equal drift parameter, the parabolic 2DEG exhibits stronger nonreciprocity than the Dirac system, although we find that a larger drift parameter does not by itself produce stronger nonreciprocity---the response is set by the interplay of drift and plasmon nonlocality. As such, the combination of strong nonlocality and experimentally accessible drift velocities of semiconductor 2DEGs identifies parabolic systems as a promising platform for tunable, nonreciprocal plasmonics.

\end{abstract}

\date{\today}
\maketitle


\textit{Introduction}---Lorentz reciprocity is a fundamental principle of classical electromagnetism stating that, in linear and time-invariant media, the electromagnetic field measured at a point $B$ due to a source at point $A$ is identical to the field at $A$ if the same source were placed at $B$ \cite{potton2004reciprocity}. In other words, the response of the system is symmetric under exchange of source and detector. This principle governs the propagation of light in the majority of conventional optical systems and its validity constrains the design of photonic devices.

On the contrary, nonreciprocal photonics---exchanging source and detector yields a different response---enables important functionalities such as optical isolation \cite{Chin2013_Nonreciprocal,Davoyan2014_Electrically}, unidirectional waveguiding \cite{Yu_2008_oneway,Bliokh18_Electric,rodriguezecharri2026nonreciprocal,Li2024_Unidirectional}, and control of radiative heat flow \cite{Hassani_2022_drifting,Liu2022_Thermal,yang2024nonreciprocal}. Nonreciprocity is conventionally achieved by breaking time-reversal symmetry through applied static magnetic fields, where the enhanced light--matter interactions associated with plasmon resonances can boost the otherwise intrinsically weak magneto-optical response of noble metals \cite{Chin2013_Nonreciprocal}. However, the need for bulky external magnets and the associated losses motivate the search for alternative, compact nanophotonic platforms that exhibit robust nonreciprocal behavior without relying on magnetic biasing or power-intensive nonlinear optical processes.

In this regard, electrically doped two-dimensional (2D) materials, and graphene in particular, have emerged as a promising platform for active plasmonics, supporting long-lived and highly confined plasmon resonances that can be tuned by injecting charge carriers~\cite{koppens2011graphene,garciadeabajo2025roadmap}. Electrostatic gating controls the carrier density and, consequently, the Fermi energy, thereby directly tuning the plasmon wavelength and confinement \cite{Chen2012_Optical,Fei2012_Gate}. However, as the plasmon wavelength is pushed toward the Fermi wavelength, finite quantum-size effects become increasingly important. These effects manifest as a pronounced wavevector dependence of the optical conductivity, rendering the electronic response intrinsically nonlocal \cite{gonccalves2016introduction,monticone2025nonlocality}. In this regime, the local Drude approximation breaks down and the full nonlocal conductivity must be retained to correctly describe plasmon propagation---
a powerful probe of this nonlocal response \cite{Torre_2015_nonlocal,Lundeberg_2017_tuning}.

\begin{figure}[!ht]
\centering
\includegraphics[width=0.35\textwidth]{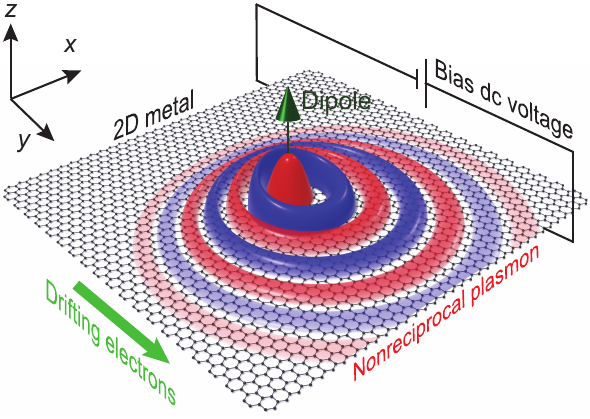}
\caption{\textbf{Schematics of a 2D metal under drift-induced nonreciprocity.} A dc voltage induces an electron drift in a 2DEG, giving rise to nonreciprocal plasmon propagation.}
\label{drift_bias_schematics}
\end{figure}

It is precisely this wavevector-dependent conductivity that, under a finite drift bias, acquires a directional asymmetry and provides a route to nonreciprocal plasmonics that circumvents the use of large applied magnetic fields \cite{Morgado_2018_DriftInduced,CorreasSerrano_2019_Nonreciprocal,Morgado_2021_Active,Morgado_2022_Directional} (Fig. \ref{drift_bias_schematics}). Experimental demonstrations of the plasmonic Fizeau effect---wherein drifting carriers drag surface plasmon polaritons, shifting their wavelength depending on the relative orientation of propagation and current flow---have established that drift-biased graphene can appreciably break the reciprocity of plasmon propagation \cite{dong2021fizeau,zhao2021efficient,dong2025currentdriven}. Theoretical studies have further shown that this nonreciprocal response is rooted in the nonlinear electrodynamics of Dirac electrons \cite{dong2021fizeau}, going beyond the simple kinematic Doppler shift of initial studies \cite{CorreasSerrano_2019_Nonreciprocal}. Such nonreciprocity can be harnessed to control the directional emission of nearby quantum light emitters \cite{eriksen2022optoelectronic,rodriguezecharri2026nonreciprocal}. 

In this work, we develop a nonlocal electrodynamic framework for drift-biased two-dimensional electron systems with Dirac and parabolic dispersions. Starting from the Boltzmann transport equation (BTE), we derive analytical expressions for the drift-dependent nonlocal conductivity tensor and use them to determine the plasmon dispersion and reflected electromagnetic Green’s tensor in the quasi-static regime, enabling us to characterize the nonreciprocal near-field response of a point dipole source. We find pronounced nonreciprocity in both Dirac and parabolic systems, with the parabolic 2DEG exhibiting a stronger response at equal drift parameter. Specifically, at fixed carrier density, the parabolic 2DEG exhibits Galilean drag, whereas the Dirac band exhibits a weaker drag that falls below the Galilean prediction. Together with the strong nonlocality and drift velocities already demonstrated in semiconductor 2DEGs, this suggests that parabolic 2DEGs may provide a promising alternative to Dirac materials for nonreciprocal plasmonics.





\textit{Conductivity of drift-biased two-dimensional electron gases}---The optical response of a two-dimensional electron gas (2DEG) to an electric field $\Eb(\rb,t)$ is encoded in the conductivity tensor $\sigma$, defined via Ohm's law $\Jb=\sigma\cdot\Eb$ from the 2D current density
\begin{equation} \label{eq:J}
    \Jb(\rb,t) = -eg\int\frac{{\rm d}^2\kb}{(2\pi)^2} \vb_\kb \fk(\rb,t) ,
\end{equation}
where $\kb$ is the in-plane electron wave vector, $\vb_\kb=\hbar^{-1}\nabla_\kb E_\kb$ is the group velocity set by dispersion relation $E_\kb$, and $\fk$ is the electronic distribution function. To find $\fk$ in the presence of a monochromatic plane-wave field $\Eb(\rb,t)=\Eb\ee^{\ii(\qb\cdot\rb-\ww t)}+{\rm c.c.}$,
we invoke the BTE in the relaxation-time approximation,
\begin{equation} \label{eq:BTE}
    \frac{{\rm d}\fk}{{\rm d}t} = \pd{\fk}{t} - \frac{e}{\hbar}\Eb\cdot\nabla_\kb\fk + \vb_\kb\cdot\nabla_\rb\fk = -\gamma\ccpar{\fk-\fk^{(0)}} ,
\end{equation}
where $\gamma$ is the phenomenological scattering rate at which the 2DEG relaxes to the equilibrium state $\fk^{(0)}$. Linearizing in $\Eb$, using the zero-temperature result $\nabla_\kb\fk^{(0)}=-(\kkb-\nabla_\kb\kF)\delta(\kF-k)$, and inserting into Eq.~\eqref{eq:J},
we identify the nonlocal conductivity tensor
\begin{equation} \label{eq:sigma_general}
    \sigma(\qb, \ww) = \frac{\ii e^2 g}{4\pi^2\hbar}\int{\rm d}^2\kb\frac{\delta(\kF-k)}{\ww+\ii\gamma-\qb\cdot\vb_\kb}\vb_\kb\otimes\ccpar{\kkb-\nabla_\kb\kF} ,
\end{equation}
where $g=g_{\rm s}g_{\rm v}$ is the spin-valley degeneracy factor and $\kF$ denotes the Fermi wave vector. The specific form of $\sigma$ follows from the dispersion relation $E_\kb$. 


For a Dirac (linear) dispersion $E_\kb^{\rm D} = \hbar v_{\rm F}^{\rm D} k$, the velocity is $\vb_\kb=\vF^{\rm D}\kkb$ (with $\nabla_\kb\kF=0$), where $\vF^{\rm D}$ is the (constant) Fermi velocity (i.e., $\vF^{\rm D}\approx c/300$ in graphene), and the carrier density $n_{2{\rm D}}$ sets the Fermi energy $E_{\rm F}^{\rm D} = 2\hbar v_{\rm F}^{\rm D}\sqrt{\pi n_{\rm 2D}/g_{\rm D}}$.

In contrast, for a parabolic dispersion $E_\kb^{\rm p}=\hbar^2k^2/(2m^*)$, the Fermi velocity is $v_{\rm F}^{\rm p}=\hbar k_{\rm F}^{\rm p}/m^*$ with $k_{\rm F}^{\rm p}=2\sqrt{\pi n_{\rm 2D}/g_{\rm p}}$ and $E_{\rm F}^{\rm p} = 2\hbar^2\pi n_{\rm 2D}/(g_{\rm p}m^*)$.

\begin{figure*}[!ht]
\centering
\includegraphics[width=0.9\textwidth]{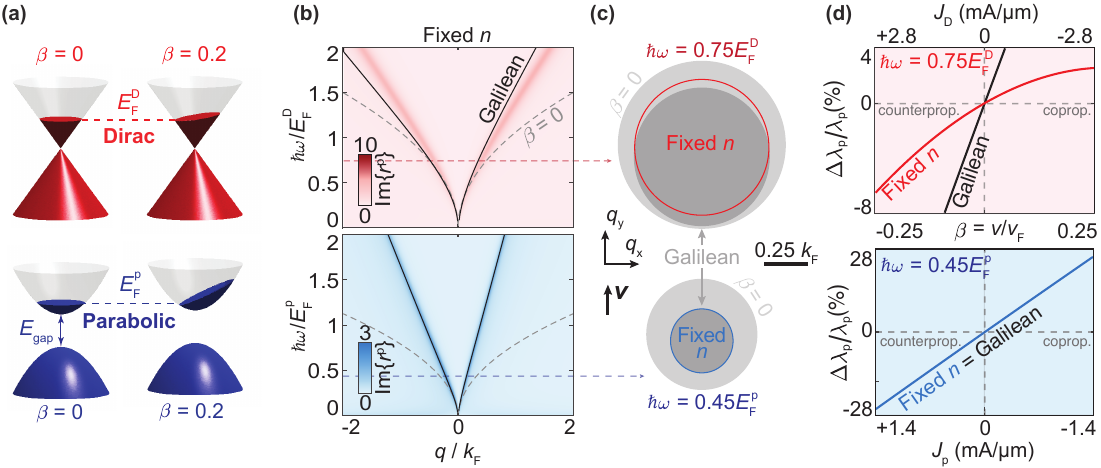}
\caption{\textbf{Drift-induced nonreciprocity in 2DEGs with Dirac and parabolic dispersions.} \textbf{(a)} Schematic band structures and corresponding Fermi surfaces in momentum space for Dirac (upper) and parabolic (lower) dispersions, without drift ($\beta=0$, left) and with drift ($\beta\neq 0$, right). The drift velocity $\vb=\beta v_{\rm F}\hat{\vb}$ biases the occupied states in $\kb$-space, breaking inversion symmetry about $\kb=0$. The parabolic Fermi sea is rigidly translated by $\kb_{\rm d}=m^*\vb/\hbar$, whereas the Dirac contour is deformed. \textbf{(b)} Loss function $\Imm\clpar{r^{\rm p}(q,\ww)}$ for Dirac (upper) and parabolic (lower) 2DEGs, for $\beta=0.2$ at fixed $n$. Solid contours show the drift-biased response; the undrifted dispersion ($\beta=0$) and Galilean dispersion are marked by dashed gray and black lines, respectively. Curves are truncated before the Landau (particle–hole) continuum boundary at $\alpha_\nu=1$. \textbf{(c)} Plasmon IFCs at $\hbar\ww=0.75E_{\rm F}^{\rm D}$ and $0.45E_{\rm F}^{\rm p}$. \textbf{(d)} Fizeau shift $\Delta\lambda_{\rm p}/\lambda_{\rm p}$ as a function of $\beta$ and the corresponding dc current density $J_\nu=en_0\beta v_{\rm F}^\nu$, evaluated using the undrifted density $n_0$, at the same photon energy.}
\label{fig1}
\end{figure*}

By isotropy, Eq.~\eqref{eq:sigma_general} yields a diagonal conductivity tensor with only two independent components parallel ($\sigma_\parallel$) and perpendicular ($\sigma_\perp$) to the optical wave vector $\qb$, 
\begin{subequations} \label{eq:sigma_nu}
    \begin{align}
        \sigma_\parallel^\nu(q,\ww) &= \frac{2}{\alpha_\nu^2}\ccpar{\frac{1}{\sqrt{1-\alpha_\nu^2}}-1}\sigma^\nu(\ww) ,  \\
        \sigma_\perp^\nu(q,\ww) &= \frac{2}{\alpha_\nu^2}\ccpar{1-\sqrt{1-\alpha_\nu^2}}\sigma^\nu(\ww) ,
    \end{align}
\end{subequations}
where $\sigma^\nu(\ww) = \ii e^2 a_\nu g_\nu E_{\rm F}^\nu/\sqpar{4\pi\hbar^2(\ww+\ii\gamma)}$ is the local Drude conductivity, with prefactors $a_{\rm D}=1$, $a_{\rm p}=2$ for 2DEGs characterized by Dirac ($\nu={\rm D}$) and parabolic ($\nu={\rm p}$) dispersion relations, respectively, and $\alpha_\nu = v_{\rm F}^\nu q/(\ww+\ii\gamma)$ encodes the nonlocal response. In the local limit $\alpha\to0$, both components coincide: $\sigma_\parallel^\nu=\sigma_\perp^\nu=\sigma^\nu(\ww)$. Eqs.~\eqref{eq:sigma_nu} assume an isotropic Fermi surface centered at $\kb=0$, corresponding to zero net current (see Fig.~\ref{fig1}a). 





A finite drift velocity $\vb$ biases the occupation of states in momentum space (Fig.~\ref{fig1}a), modifying the equilibrium distribution to the skewed Fermi-Dirac function
\begin{equation} \label{eq:f0_drift}
    \fk^{(0)} = \left[\ee^{(E_\kb-\mu-\hbar\vb\cdot\kb)/\kB T}+1\right]^{-1}
\end{equation}
at chemical potential $\mu$ and finite temperature $T$. At zero temperature, this reduces to $\fk^{(0)}=\Theta(\tilde{k}_{\rm F}-k)$, with an angle-dependent Fermi wave vector $\tilde{k}_{\rm F}(\kkb)$. The drift also breaks the inversion symmetry $\sigma(q)=\sigma(-q)$, enabling nonreciprocal plasmon propagation, as we now show by evaluating Eq.~\eqref{eq:sigma_general} with the drifted distribution for each dispersion characterized by the dimensionless drift parameter $\beta\equiv v/v_{\rm F}^\nu$ (the ratio of drift speed $v=|\vb|$ to the undrifted Fermi velocity).


For a Dirac system $\vb=\beta v_{\rm F}^{\rm D}\hat{\bf v}$, the angle-dependent Fermi wave vector is
\begin{equation}
    \tilde{k}_{\rm F} = \frac{E_{\rm F}^{\rm D}}{\hbar(v_{\rm F}^{\rm D}-\vb\cdot\kkb)}.
\end{equation}
Inserting $\vb_\kb=v_{\rm F}^{\rm D}\kkb$ and $\nabla_\kb\tilde{k}_{\rm F} = \hbar \tilde{k}_{\rm F}^2[\vb-(\vb\cdot\kkb)\kkb]/(k E_{\rm F}^{\rm D})$ into Eq.~\eqref{eq:sigma_general} and evaluating the radial integral yields
\begin{equation}
    \sigma^{\rm D}(\qb,\ww) = \frac{\sigma^{\rm D}(\ww)}{\pi}\int_0^{2\pi}{\rm d}\varphi_\kb \frac{v_{\rm F}^{\rm D}\kkb\otimes(v_{\rm F}^{\rm D}\kkb-\vb)}{(1-\alpha_{\rm D}\hat{\qb}\cdot\kkb)(v_{\rm F}^{\rm D}-\vb\cdot\kkb)^2} .
\end{equation}
For the special case of propagation parallel to the drift, i.e., $\hat{\qb}=\hat{\vb}$, the angular integral evaluates analytically~\cite{rodriguezecharri2026nonreciprocal}, yielding a diagonal tensor with components
\begin{subequations} \label{eq:sigma_D_parallel}
    \begin{align}
        \sigma_\parallel^{\rm D}(q,\ww) &= \frac{2\sigma^{\rm D}(\ww)}{(\alpha_{\rm D}-\beta)^2}\ccpar{\frac{1-\alpha_{\rm D}\beta}{\sqrt{1-\alpha_{\rm D}^2}}-\sqrt{1-\beta^2}} ,  \\
        \sigma_\perp^{\rm D}(q,\ww) &= \frac{2\sigma^{\rm D}(\ww)}{(\alpha_{\rm D}-\beta)^2}\ccpar{\frac{1-\alpha_{\rm D}\beta}{\sqrt{1-\beta^2}}-\sqrt{1-\alpha_{\rm D}^2}} .
    \end{align}
\end{subequations}
In the limit $\beta\to0$, Eqs.~\eqref{eq:sigma_D_parallel} reduce exactly to Eqs.~\eqref{eq:sigma_nu}.

For a parabolic system, the skewed Fermi surface is parametrized as $\tilde{k}_{\rm F} = m^*\tilde{v}_{\rm F}^{\rm p}/\hbar$, where
\begin{equation} \label{eq:vFp_tilde}
    \tilde{v}_{\rm F}^{\rm p} = \vb\cdot\kkb+\sqrt{(\vb\cdot\kkb)^2+2E_{\rm F}^{\rm p}/m^*}
\end{equation}
is the angle-dependent Fermi velocity for a carrier density corresponding to the Fermi energy $E_{\rm F}^{\rm p}$ in the absence of drift current. Inserting $\vb_\kb=\hbar\kb/m^*$ and 
$\nabla_\kb\tilde{k}_{\rm F}=(\tilde{k}_{\rm F}/k)\sqpar{\vb-(\vb\cdot\kkb)\kkb}/\ccpar{\tilde{v}_{\rm F}^{\rm p}-\vb\cdot\kkb}$ 
into Eq.~\eqref{eq:sigma_general} and evaluating the $k$-integral, we obtain
\begin{equation}
    \sigma^{\rm p}(\qb,\ww) = \frac{\sigma^{\rm p}(\ww)}{\pi}\int_0^{2\pi}{\rm d}\varphi_\kb\frac{m^*(\tilde{v}_{\rm F}^{\rm p})^2/(2E_{\rm F}^{\rm p})\kkb\otimes(\tilde{v}_{\rm F}^{\rm p}\kkb-\vb)}{(1-\tilde{\alpha}_{\rm p}\hat{\qb}\cdot\kkb)(\tilde{v}_{\rm F}^{\rm p}-\vb\cdot\kkb)} ,
\end{equation}
where $\tilde{\alpha}_{\rm p}=\tilde{v}_{\rm F}^{\rm p} q/(\ww+\ii\gamma)$.

For optical modes propagating parallel to the drift, the conductivity tensor evaluates to
\begin{subequations} \label{eq:sigma_p_parallel}
    \begin{align} 
        \sigma_\parallel^{\rm p}(\qb,\ww) &= \frac{2\sigma^{\rm p}(\ww)}{\alpha_{\rm p}^2}\frac{1-\alpha_{\rm p}\beta-\sqrt{1-\alpha_{\rm p}^2-2\alpha_{\rm p}\beta}}{\sqrt{1-\alpha_{\rm p}^2-2\alpha_{\rm p}\beta}} ,  \\
        \sigma_\perp^{\rm p}(\qb,\ww) &= \frac{2\sigma^{\rm p}(\ww)}{\alpha_{\rm p}^2}\ccpar{1-\alpha_{\rm p}\beta-\sqrt{1-\alpha_{\rm p}^2-2\alpha_{\rm p}\beta}}
    \end{align}
\end{subequations}
where $\alpha_{\rm p}\equiv v_{\rm F}^{\rm p}q/(\ww+\ii\gamma)$ is defined using the undrifted Fermi velocity $v_{\rm F}^{\rm p}$, and $\beta\equiv v/v_{\rm F}^{\rm p}$. Setting $\beta\to0$ recovers Eq.~\eqref{eq:sigma_nu}. Since $q$ is a signed wave vector component, both Eqs.~\eqref{eq:sigma_D_parallel} and \eqref{eq:sigma_p_parallel} yield $\sigma_\parallel(q,\ww)\neq\sigma_\parallel(-q,\ww)$ for $\beta\neq 0$. Eqs. \eqref{eq:sigma_D_parallel} and \eqref{eq:sigma_p_parallel} are found for $\qb\parallel\vb$. Arbitrary directions require a general angle $\theta$ between $\qb$ and $\vb$, which we find by numerical evaluation of the same angular integral. The result for both dispersions is given in the Supplemental Material~\cite{SM}.


Equations~\eqref{eq:sigma_D_parallel} and~\eqref{eq:sigma_p_parallel} are written for fixed $\mu$, with $\beta\equiv v/v_{\rm F}^\nu$ referred throughout to the undrifted Fermi velocity.  If the carrier density is held fixed, conservation of the occupied $\kb$-space area requires
\begin{subequations} \label{eq:EF_renorm}
    \begin{align}
        \tilde{E}_{\rm F}^{\rm D} &= \left(1-\beta^2\right)^{3/4} E_{\rm F}^{\rm D} ,  \\
        \tilde{E}_{\rm F}^{\rm p} &= \left(1-\beta^2\right) E_{\rm F}^{\rm p} ,
    \end{align}
\end{subequations}
where the parabolic expression follows from $n/n_0=\mu/\mu_0+\beta^2$ when $\beta$ is defined using the undrifted $v_{\rm F}^{\rm p}$. We refer to these two idealized limits as fixed $\mu$ and fixed $n$, respectively. Note that, if the drift parameter $\tilde{\beta}$ is referenced to $\tilde{E}_{\rm F}^{\rm p}$ rather than to the undrifted $E_{\rm F}^{\rm p}$, the alternative expression $\tilde{E}_{\rm F}^{\rm p}/E_{\rm F}^{\rm p}=(1+\tilde{\beta}^2)^{-1}$ should be used.

Two regimes are excluded from this analysis. If the drift velocity exceeds the plasmon phase velocity, drifting carriers can resonantly amplify rather than damp plasmons---a 2D analogue of Cherenkov radiation---leading to an instability not captured here. However, experimentally accessible drift velocities typically remain well below this threshold~\cite{dong2021fizeau}. Additionally, drift modifies the threshold for interband transitions by shifting the occupied states away from the Dirac point or band minimum, an effect neglected here as we focus on the heavily doped intraband regime $\EF\gg\hbar\ww$.

\textit{Nonreciprocal plasmon dispersion under drift bias}---The impact of drift current on the plasmonic response is visualized through the loss function $\Imm\clpar{r^{\rm p}(q,\ww)}$, where $r^{\rm p}$ is the p-polarized Fresnel reflection coefficient for a 2D sheet sandwiched between two dielectric half-spaces (see \cite{SM} for the full expression). For the Dirac case we take hBN-encapsulated graphene, with $n = 7\times10^{12}$\,cm$^{-2}$, $v_F^{\rm D} = 10^6$\,m/s, $g_{\rm D} = 4$ (spin $\times$ valley), $\epsilon_{\rm r} = 3.9$, and $\gamma = 37$\,meV$/\hbar$. For the parabolic case we take an InGaAs/InAlAs 2DEG with $n = 3\times10^{12}$~cm$^{-2}$, representative of the densities reported for such heterojunctions~\cite{gueissaz1991high}, together with $m^* = 0.043\,m_e$ and $\epsilon_\infty = 11.6$ for In$_{0.53}$Ga$_{0.47}$As~\cite{adachi1982material}, $g_{\rm p} = 2$ (spin only), and $\gamma = 13$~meV$/\hbar$, chosen so that $\gamma/\omega$ matches the graphene case at the frequencies considered below. These give $\EF = 309$ and $167$\,meV and $v_F^{\rm p} = 1.17\times10^6$\,m/s. Moreover, we use $\beta = 0.2$ for both dispersions. For graphene this is of the order of the drift velocities $v \approx 0.15\,v_F^{\rm D}$ demonstrated in hBN-encapsulated samples~\cite{dong2021fizeau}. For InGaAs it corresponds to $v_{\rm d} = 2.3\times10^5$\,m/s, just below the $2.5\times10^5$\,m/s measured in InGaAs quantum wells~\cite{pozela2011electron} (see \cite{SM} for a more detailed discussion on experimentally achievable values of $\beta$).

Fig.~\ref{fig1}b shows $\Imm\clpar{r^{\rm p}(q,\ww)}$ for both Dirac and parabolic 2DEGs at fixed $n$ (the fixed-$\mu$ limit is shown in \cite{SM}), using $\beta = 0.2$ up to $\hbar\ww=2E_F$, the onset of the interband regime in graphene \cite{gonccalves2016introduction}. For the 2D parabolic material, this is justified by assuming that it exhibits some band gap, such that the $q=0$ threshold for interband would be $\hbar\ww>E_{\rm gap}+\EF$, i.e., from the top of the valence band to the lowest unoccupied (undrifted) state in the conduction band. Peaks in $\Imm\clpar{r^{\rm p}(q,\ww)}$ indicate the dispersion relation, different for the forward $(q>0)$ and backward $(q<0)$ plasmon branches. The Dirac and parabolic cases exhibit distinct asymmetries rooted in their different drifted Fermi contours, with the parabolic almost identical to the Galilean prediction. This is more clearly shown in Fig.~\ref{fig1}c, where peaks in $\Imm\clpar{r^{\rm p}(q,\ww)}$ at fixed energy are mapped in $\qb$-space to give the plasmon isofrequency contours (IFCs), clearly different for Dirac and parabolic. For the Dirac dispersion, $|\vb_\kb|=v_{\rm F}^{\rm D}$ remains constant but the radial Fermi wave vector becomes angle-dependent, so the contour is deformed rather than rigidly translated. For the parabolic dispersion, completing the square in $E_\kb^{\rm p}-\hbar\vb\cdot\kb$ shows that the occupied region is always a circle centered at $\kb_{\rm d}=m^*\vb/\hbar$. At fixed $n$ its radius is unchanged, while at fixed $\mu$ \cite{SM} its radius increases, with the parabolic Fermi sea being translated and enlarged, not angularly distorted.

Fig.~\ref{fig1}d shows the Fizeau shift $\Delta\lambda_{\rm p}/\lambda_{\rm p}$ of the plasmon wavelength induced by the drift bias---where $\Delta\lambda_{\rm p} = \lambda_{\rm p}^\pm - \lambda_{\rm p}$ is the difference between backward- or forward propagating plasmon wavelengths and the undrifted value $\lambda_{\rm p} = 2\pi/q_{\rm p}$ at each frequency---normalized to the undrifted value. The Galilean transformation predicts a Doppler kinematic shift set by the ratio of the drift speed to plasmon group velocity (black line in Fig.~\ref{fig1}d). The full BTE result for Dirac systems falls below this kinematic prediction, in agreement with previous results \cite{dong2021fizeau}. Moreover, the shift is asymmetric about zero drift. At fixed $n$ the larger shift falls on the counter-propagating side instead, which is the ordering observed experimentally. In fact, inserting the parameters used in the experiment \cite{dong2021fizeau}, $n=2.9\times10^{12}$~cm$^{-2}$, $\epsilon_{\rm r}=1.6$, $\hbar\ww=890$~cm$^{-1}$, and $\beta=0.15$, we obtain $-2.49\%$ counter-propagating and $+0.34\%$ co-propagating, in agreement with $-2.5\%$ and a small positive shift reported in \cite{dong2021fizeau}. The counter-propagating shift and asymmetry are well reproduced, while the co-propagating branch is slightly underestimated. This is expected since Ref.~\cite{dong2021fizeau} attributes part of their measured response to higher-order nonlinear electrodynamics beyond the displaced-Fermi-sea model used here. At fixed $\mu$ \cite{SM}, co-propagation produces a larger absolute shift than counter-propagation at equal $|\beta|$ owing to the stronger modification of the optical conductivity when drift and wave travel in the same direction. We also show in \cite{SM} how the response evolves continuously between the fixed $\mu$ or fixed $n$ limits.

On the other hand, a parabolic 2DEG behaves differently: at fixed $n$ its drifted Fermi sea is the undrifted one translated rigidly by $m^*\vb/\hbar$ (Fig.~\ref{fig1}c), so the BTE result lies on the Galilean line (Fig.~\ref{fig1}d). At fixed $\mu$ \cite{SM} the displaced Fermi sea encloses $(1+\beta^2)$ times the equilibrium carriers and the two no longer coincide. The Galilean transformation therefore overestimates the drag for Dirac carriers but is exact for a parabolic band at fixed $n$. Thus, the parabolic 2DEG exhibits a larger Fizeau shift than the Dirac one at equal $\beta$. At $\beta=0.2$ the drag part of the shift is $+3.7\%$ for graphene and $+20.4\%$ for InGaAs, partially due to the stronger nonlocality of the InGaAs plasmon ($\alpha_{\rm p}=0.77$ against $\alpha_{\rm D}=0.53$). 

\begin{figure}[!ht]
\centering
\includegraphics[width=0.35\textwidth]{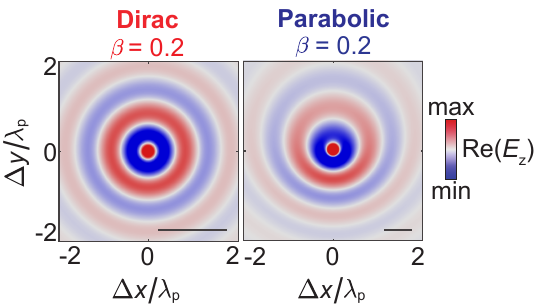}
\caption{\textbf{Electric near-field distributions for 2DEGs with Dirac and parabolic dispersions under drift-induced nonreciprocity.} Real-space maps of the electric field $\text{Re}(E_z)$ generated by a $z$-oriented point dipole at height $z_0=10$~nm above Dirac (left) and parabolic (right) 2DEGs, at $\beta=0.2$, computed from the BTE at fixed $n$. The photon energies are $\hbar\ww_0=0.75E_{\rm F}^{\rm D}$ for the Dirac panel and $0.45E_{\rm F}^{\rm p}$ for the parabolic panel. Scale bars: 50 nm}
\label{fig2}
\end{figure}

\textit{Nonreciprocal emission of light under drift bias}---The nonreciprocal plasmon dispersion has direct consequences for the emission of nearby quantum light sources. A point dipole placed at height $z_0$ above the 2DEG couples to the plasmonic modes of the sheet, and its emission pattern is governed by the reflected part of the electromagnetic Green's tensor $\Gm^{\rm ref}(\rb,\rb',\ww)$, obtained by integrating the Fresnel coefficients over in-plane wave vectors \cite{SM}. Figure \ref{fig2} shows the $z$-component real-space field distribution $E_z$ for a $z$-dipole oriented normally ($z_0=10$~nm) to both Dirac (left) and parabolic (right) 2DEGs, for which the relevant quasi-static Green-tensor component is $\Gm_{zz}^{\rm ref}$. We use $\hbar\ww_0=0.75E_{\rm F}^{\rm D}$ for graphene and $0.45E_{\rm F}^{\rm p}$ for InGaAs, matching the frequency slices of Figs.~\ref{fig1}c and \ref{fig1}d, both at fixed $n$ with $\beta=0.2$. Drift breaks the symmetry of the plasmon wavefronts for both about the dipole, yielding different propagation parallel or antiparallel to the drift bias, a direct spatial manifestation of the nonreciprocal dispersion. Because the drifted Dirac contour is deformed rather than translated, the full BTE treatment modifies both the wavelength and the spatial profile of the launched plasmons relative to the Galilean shift. For the parabolic 2DEG at fixed $n$, by contrast, the map calculated from the BTE conductivity reproduces the Galilean boost up to a small finite-$\gamma$ artifact. The corresponding maps at fixed $\mu$, which do depart from the Galilean result, are given in \cite{SM}.


The different nonreciprocal responses originate from the interplay between two factors. First, for a parabolic band at fixed $n$, the Fermi sea undergoes a rigid translation, and the BTE conductivity equals the equilibrium nonlocal conductivity at the Doppler-shifted frequency \cite{SM}. In contrast, a Dirac band is not Galilean invariant: drift deforms its Fermi contour, reducing the BTE drag relative to the Galilean prediction. Second, the magnitude of the nonreciprocal response depends strongly on the nonlocality parameter $\alpha$.
Expanding the fixed-$\mu$ conductivities in Eqs.~\eqref{eq:sigma_D_parallel} and \eqref{eq:sigma_p_parallel} to leading order in $\beta$ gives $\Am^{\rm D} = 4\beta[(2-\alpha^2)/\sqrt{1-\alpha^2}-2]/\alpha^3$ and $\Am^{\rm p} = 4\beta\alpha/(1-\alpha^2)^{3/2}$, where  $\Am^\nu \equiv |\sigma_\parallel^\nu(q,\ww)-\sigma_\parallel^\nu(-q,\ww)|/|\sigma^\nu(\ww)|$  measures the directional asymmetry normalized to the local conductivity. The parabolic asymmetry exceeds the Dirac one for a given $\alpha$, and their ratio approaches $(1-\alpha^2)^{-1}$ as $\alpha\to1$. For the parameters used here, the InGaAs plasmon is also more nonlocal than the graphene plasmon \cite{SM}. These two ingredients---the exact Galilean parabolic shift of a 2DEG and the stronger nonlocality accessible in InGaAs---account for its more nonreciprocal parabolic response. A heavier band can support larger $\beta$, but not necessarily stronger nonreciprocity. For example, AlGaN/GaN 2DEGs with $n\simeq2\times10^{13}$~cm\textsuperscript{-2} and $m^*=0.2m_e$ reach $v_{\rm d}=3.1\times10^5$~m/s~\cite{barker2005high}, corresponding to $\beta\simeq0.5$, yet the predicted asymmetry remains smaller than for InGaAs at $\beta=0.2$ because the GaN plasmon is less nonlocal; a quantitative comparison for GaN is complicated by its polar-optical-phonon response and strongly nonlinear high-field transport \cite{SM}. In any case, the effect is controlled by both drift and nonlocality, not by $\beta$ alone. The parabolic band also offers a tuning knob absent in the Dirac case: since $\beta\propto n^{-1/2}$ at fixed drift velocity, gating alone tunes the nonreciprocity, while $\beta$ is independent of $n$ for a Dirac band \cite{SM}.

\textit{Conclusion}---Our closed-form expressions for the nonlocal conductivity tensor, derived from the BTE under finite drift bias, show a splitting of the forward and backward plasmon branches for both dispersions. The resulting nonreciprocities differ because the two bands transform differently under drift. For a parabolic band at fixed $n$, the drifted Fermi sea is a rigid translation in momentum and the BTE response is exactly Galilean. For a Dirac band, the drifted contour is deformed and the drag is suppressed relative to the Galilean prediction. In real space, this nonreciprocity leads to preferential plasmon launching along the drift direction. For the material parameters studied here, the parabolic InGaAs 2DEG exhibits the stronger response at equal drift parameter, aided by its stronger nonlocality, while the large drift velocities accessible in semiconductor heterostructures provide an additional practical advantage. However, the two ingredients are not interchangeable: a heavier band such as AlGaN/GaN reaches more than twice the drift parameter yet a smaller predicted asymmetry, since its plasmon is considerably less nonlocal. Overall, drift-biased 2D electron systems provide a tunable, magnet-free platform for nonreciprocal nanophotonics and a useful setting for isolating how band dispersion and nonlocality control plasmonic drag.

\textit{Acknowledgments}---G.~{\'A}.-P. acknowledges support from the European Union (Marie Skłodowska-Curie Actions, grant agreement No. 101209198). The Center for Polariton-driven Light--Matter Interactions (POLIMA) is funded by the Danish National Research Foundation (Project No.~DNRF165).


\textit{Data availability}---The data that support the findings of this article are available from the authors upon reasonable request.

\bibliographystyle{apsrev4-2}
\bibliography{refs}

@article{adachi1982material,
    author = {Adachi, Sadao},
    title = {Material parameters of In1−xGaxAsyP1−y and related binaries},
    journal = {J. Appl. Phys.},
    volume = {53},
    number = {12},
    pages = {8775-8792},
    year = {1982},
    month = {12},
    issn = {0021-8979},
    doi = {10.1063/1.330480},
    url = {https://doi.org/10.1063/1.330480},
}

@article{potton2004reciprocity,
  title     = {Reciprocity in optics},
  author    = {Richard J. Potton},
  journal   = {Rep.\ Prog.\ Phys.},
  volume    = {67},
  number    = {5},
  pages     = {717},
  year      = {2004},
  publisher = {IOP Publishing},
  doi       = {10.1088/0034-4885/67/5/R03}
}

@article{Yu_2008_oneway,
  title = {One-Way Electromagnetic Waveguide Formed at the Interface between a Plasmonic Metal under a Static Magnetic Field and a Photonic Crystal},
  author = {Yu, Zongfu and Veronis, Georgios and Wang, Zheng and Fan, Shanhui},
  journal = {Phys. Rev. Lett.},
  volume = {100},
  issue = {2},
  pages = {023902},
  numpages = {4},
  year = {2008},
  month = {Jan},
  publisher = {American Physical Society},
  doi = {10.1103/PhysRevLett.100.023902},
  url = {https://link.aps.org/doi/10.1103/PhysRevLett.100.023902}
}

@article{koppens2011graphene,
  title     = {Graphene plasmonics: a platform for strong light--matter interactions},
  author    = {Frank H. L. Koppens and Darrick E. Chang and F. J. {Garc\'{\i}a de Abajo}},
  journal   = {Nano Lett.},
  volume    = {11},
  number    = {8},
  pages     = {3370--3377},
  year      = {2011},
  publisher = {ACS Publications},
  doi       = {10.1021/nl201771h}
}

@Article{Fei2012_Gate,
author={Fei, Z.
and Rodin, A. S.
and Andreev, G. O.
and Bao, W.
and McLeod, A. S.
and Wagner, M.
and Zhang, L. M.
and Zhao, Z.
and Thiemens, M.
and Dominguez, G.
and Fogler, M. M.
and Neto, A. H. Castro
and Lau, C. N.
and Keilmann, F.
and Basov, D. N.},
title={Gate-tuning of graphene plasmons revealed by infrared nano-imaging},
journal={Nature},
year={2012},
month={Jul},
day={01},
volume={487},
number={7405},
pages={82-85},
issn={1476-4687},
doi={10.1038/nature11253},
url={https://doi.org/10.1038/nature11253}
}

@Article{Chen2012_Optical,
author={Chen, Jianing
and Badioli, Michela
and Alonso-Gonz{\'a}lez, Pablo
and Thongrattanasiri, Sukosin
and Huth, Florian
and Osmond, Johann
and Spasenovi{\'{c}}, Marko
and Centeno, Alba
and Pesquera, Amaia
and Godignon, Philippe
and Zurutuza Elorza, Amaia
and Camara, Nicolas
and de Abajo, F. Javier Garc{\'i}a
and Hillenbrand, Rainer
and Koppens, Frank H. L.},
title={Optical nano-imaging of gate-tunable graphene plasmons},
journal={Nature},
year={2012},
month={Jul},
day={01},
volume={487},
number={7405},
pages={77-81},
issn={1476-4687},
doi={10.1038/nature11254},
url={https://doi.org/10.1038/nature11254}
}

@Article{Chin2013_Nonreciprocal,
    author={Chin, Jessie Yao
    and Steinle, Tobias
    and Wehlus, Thomas
    and Dregely, Daniel
    and Weiss, Thomas
    and Belotelov, Vladimir I.
    and Stritzker, Bernd
    and Giessen, Harald},
    title={Nonreciprocal plasmonics enables giant enhancement of thin-film Faraday rotation},
    journal={Nat. Commun.},
    year={2013},
    month={Mar},
    day={19},
    volume={4},
    number={1},
    pages={1599},
    issn={2041-1723},
    doi={10.1038/ncomms2609},
    url={https://doi.org/10.1038/ncomms2609}
}

@Article{Davoyan2014_Electrically,
    author={Davoyan, Artur
    and Engheta, Nader},
    title={Electrically controlled one-way photon flow in plasmonic nanostructures},
    journal={Nat. Commun.},
    year={2014},
    month={Nov},
    day={06},
    volume={5},
    number={1},
    pages={5250},
    issn={2041-1723},
    doi={10.1038/ncomms6250},
    url={https://doi.org/10.1038/ncomms6250}
}

@article{Torre_2015_nonlocal,
  title = {Nonlocal transport and the hydrodynamic shear viscosity in graphene},
  author = {Torre, Iacopo and Tomadin, Andrea and Geim, Andre K. and Polini, Marco},
  journal = {Phys. Rev. B},
  volume = {92},
  issue = {16},
  pages = {165433},
  numpages = {11},
  year = {2015},
  month = {Oct},
  publisher = {American Physical Society},
  doi = {10.1103/PhysRevB.92.165433},
  url = {https://link.aps.org/doi/10.1103/PhysRevB.92.165433}
}

@book{gonccalves2016introduction,
  title     = {An introduction to graphene plasmonics},
  author    = {Paulo Andr{\'e} Dias Gon{\c{c}}alves and Nuno M. R. Peres},
  year      = {2016},
  publisher = {World Scientific},
  url = {https://www.worldscientific.com/worldscibooks/10.1142/9948?srsltid=AfmBOorg6Xy0-uOA4lLXCMalol8PSo1VDdp0ISsQrUJTgOb2olPm0waR#t=aboutBook}
}

@article{Lundeberg_2017_tuning,
    author = {Mark B. Lundeberg  and Yuanda Gao  and Reza Asgari  and Cheng Tan  and Ben Van Duppen  and Marta Autore  and Pablo Alonso-González  and Achim Woessner  and Kenji Watanabe  and Takashi Taniguchi  and Rainer Hillenbrand  and James Hone  and Marco Polini  and Frank H. L. Koppens },
    title = {Tuning quantum nonlocal effects in graphene plasmonics},
    journal = {Science},
    volume = {357},
    number = {6347},
    pages = {187-191},
    year = {2017},
    doi = {10.1126/science.aan2735},
    URL = {https://www.science.org/doi/abs/10.1126/science.aan2735},
}

@article{Morgado_2018_DriftInduced,
    author = {Morgado, Tiago A. and Silveirinha, Mário G.},
    title = {Drift-Induced Unidirectional Graphene Plasmons},
    journal = {ACS Photonics},
    volume = {5},
    number = {11},
    pages = {4253-4258},
    year = {2018},
    month = {10},
    issn = {2330-4022},
    doi = {10.1021/acsphotonics.8b00987},
    url = {https://doi.org/10.1021/acsphotonics.8b00987},
}

@article{Bliokh18_Electric,
    author = {K. Y. Bliokh and F. J. Rodr\'{i}guez-Fortu\~{n}o and A. Y. Bekshaev and Y. S. Kivshar and F. Nori},
    journal = {Opt. Lett.},
    number = {5},
    pages = {963--966},
    publisher = {Optica Publishing Group},
    title = {Electric-current-induced unidirectional propagation of surface plasmon-polaritons},
    volume = {43},
    month = {Mar},
    year = {2018},
    url = {https://opg.optica.org/ol/abstract.cfm?URI=ol-43-5-963},
    doi = {10.1364/OL.43.000963},
}

@article{CorreasSerrano_2019_Nonreciprocal,
  title = {Nonreciprocal and collimated surface plasmons in drift-biased graphene metasurfaces},
  author = {Correas-Serrano, D. and Gomez-Diaz, J. S.},
  journal = {Phys. Rev. B},
  volume = {100},
  issue = {8},
  pages = {081410(R)},
  numpages = {6},
  year = {2019},
  month = {Aug},
  publisher = {American Physical Society},
  doi = {10.1103/PhysRevB.100.081410},
  url = {https://link.aps.org/doi/10.1103/PhysRevB.100.081410}
}

@article{dong2021fizeau,
  title     = {Fizeau drag in graphene plasmonics},
  author    = {Y. Dong and L. Xiong and I. Y. Phinney and Z. Sun and R. Jing and A. S. McLeod and S. Zhang and S. Liu and F. L. Ruta and H. Gao and Z. Dong and R. Pan and J. H. Edgar and P. Jarillo-Herrero and L. S. Levitov and A. J. Millis and M. M. Fogler and D. A. Bandurin and D. N. Basov},
  journal   = {Nature},
  volume    = {594},
  number    = {7864},
  pages     = {513--516},
  year      = {2021},
  publisher = {Nature Publishing Group UK London},
  doi       = {10.1038/s41586-021-03640-x}
}

@article{zhao2021efficient,
  title     = {Efficient Fizeau drag from Dirac electrons in monolayer graphene},
  author    = {Wenyu Zhao and Sihan Zhao and Hongyuan Li and Sheng Wang and Shaoxin Wang and M. {Iqbal Bakti Utama} and Salman Kahn and Yue Jiang and Xiao Xiao and SeokJae Yoo and Kenji Watanabe and Takashi Taniguchi and Alex Zettl and Feng Wang},
  journal   = {Nature},
  volume    = {594},
  number    = {7864},
  pages     = {517--521},
  year      = {2021},
  publisher = {Nature Publishing Group UK London},
  doi       = {10.1038/s41586-021-03574-4}
}

@article{Morgado_2021_Active,
    author = {Morgado, Tiago
A. and Silveirinha, Mário G.},
    title = {Active Graphene Plasmonics with a Drift-Current Bias},
    journal = {ACS Photonics},
    volume = {8},
    number = {4},
    pages = {1129-1136},
    year = {2021},
    month = {04},
    issn = {2330-4022},
    doi = {10.1021/acsphotonics.0c01890},
    url = {https://doi.org/10.1021/acsphotonics.0c01890},
}

@article{Morgado_2022_Directional,
url = {https://doi.org/10.1515/nanoph-2022-0451},
title = {Directional dependence of the plasmonic gain and nonreciprocity in drift-current biased graphene},
title = {},
author = {Tiago A. Morgado and Mário G. Silveirinha},
pages = {4929--4936},
volume = {11},
number = {21},
journal = {Nanophotonics},
doi = {doi:10.1515/nanoph-2022-0451},
year = {2022},
}

@article{Hassani_2022_drifting,
    author = {Hassani Gangaraj, S. Ali and Monticone, Francesco},
    title = {Drifting Electrons: Nonreciprocal Plasmonics and Thermal Photonics},
    journal = {ACS Photonics},
    volume = {9},
    number = {3},
    pages = {806-819},
    year = {2022},
    doi = {10.1021/acsphotonics.1c01294},
    URL = {https://doi.org/10.1021/acsphotonics.1c01294}
}

@Article{Liu2022_Thermal,
    author={Liu, Tianji
    and Guo, Cheng
    and Li, Wei
    and Fan, Shanhui},
    title={Thermal photonics with broken symmetries},
    journal={eLight},
    year={2022},
    month={Dec},
    day={02},
    volume={2},
    number={1},
    pages={25},
    issn={2662-8643},
    doi={10.1186/s43593-022-00025-z},
    url={https://doi.org/10.1186/s43593-022-00025-z}
}

@article{eriksen2022optoelectronic,
  title     = {Optoelectronic Control of Atomic Bistability with Graphene},
  author    = {Mikkel Have Eriksen and Jakob E. Olsen and Christian Wolff and Joel D. Cox},
  journal   = {Phys.\ Rev.\ Lett.},
  volume    = {129},
  issue     = {25},
  pages     = {253602},
  year      = {2022},
  publisher = {APS},
  doi       = {10.1103/PhysRevLett.129.253602}
}

@article{yang2024nonreciprocal,
  title     = {Nonreciprocal thermal photonics},
  author    = {Shuihua Yang and Mengqi Liu and Changying Zhao and Shanhui Fan and Cheng-Wei Qiu},
  journal   = {Nat.\ Photonics},
  volume    = {18},
  number    = {5},
  pages     = {412--424},
  year      = {2024},
  publisher = {Nature Publishing Group UK London},
  doi       = {10.1038/s41566-024-01409-y}
}

@Article{Li2024_Unidirectional,
    author={Li, Shiqing
    and Tsakmakidis, Kosmas L.
    and Jiang, Tao
    and Shen, Qian
    and Zhang, Hang
    and Yan, Jinhua
    and Sun, Shulin
    and Shen, Linfang},
    title={Unidirectional guided-wave-driven metasurfaces for arbitrary wavefront control},
    journal={Nat. Commun.},
    year={2024},
    month={Jul},
    day={16},
    volume={15},
    number={1},
    pages={5992},
    issn={2041-1723},
    doi={10.1038/s41467-024-50287-z},
    url={https://doi.org/10.1038/s41467-024-50287-z}
}

@article{dong2025currentdriven,
  title     = {Current-driven nonequilibrium electrodynamics in graphene revealed by nano-infrared imaging},
  author    = {Y. Dong and Z. Sun and I. Y. Phinney and D. Sun and T. I. Andersen and L. Xiong and Y. Shao and S. Zhang and Andrey Rikhter and S. Liu and P. Jarillo-Herrero and P. Kim and C. R. Dean and A. J. Millis and M. M. Fogler and D. A. Bandurin and D. N. Basov},
  journal   = {Nat.\ Commun.},
  volume    = {16},
  number    = {1},
  pages     = {3861},
  year      = {2025},
  publisher = {Nature Publishing Group UK London},
  doi       = {10.1038/s41467-025-58953-6}
}

@article{monticone2025nonlocality,
  title     = {Nonlocality in photonic materials and metamaterials: roadmap},
  author    = {Francesco Monticone and N. Asger Mortensen and Antonio I. Fern\'{a}ndez-Dom\'{i}nguez and Yu Luo and Xuezhi Zheng and Christos Tserkezis and Jacob B. Khurgin and Tigran V. Shahbazyan and Andr\'{e} J. Chaves and Nuno M. R. Peres and Gino Wegner and Kurt Busch and Huatian Hu and Fabio Della Sala and Pu Zhang and Cristian Cirac\`{i} and Javier Aizpurua and Antton Babaze and Andrei G. Borisov and Xue-Wen Chen and Thomas Christensen and Wei Yan and Yi Yang and Ulrich Hohenester and Lorenz Huber and Martijn Wubs and Simone De Liberato and P. A. D. Gon\c{c}alves and F. Javier {Garc\'{\i}a de Abajo} and Ortwin Hess and Illya Tarasenko and Joel D. Cox and Line Jelver and Eduardo J. C. Dias and Miguel S\'{a}nchez S\'{a}nchez and Dionisios Margetis and Guillermo G\'{o}mez-Santos and Igor M. Vasilevskiy and Tobias Stauber and Sergei Tretyakov and Constantin Simovski and Samaneh Pakniyat and J. Sebasti\'{a}n G\'{o}mez-D\'{i}az and Igor V. Bondarev and Svend-Age Biehs and Alexandra Boltasseva and Vladimir M. Shalaev and Alexey V. Krasavin and Anatoly V. Zayats and Andrea Al\`{u} and Jung-Hwan Song and Mark L. Brongersma and Uriel Levy and Olivia Y. Long and Cheng Guo and Shanhui Fan and Sergey I. Bozhevolnyi and Adam Overvig and Filipa R. Prud\^{e}ncio and M\'{a}rio G. Silveirinha and S. Ali Hassani Gangaraj and Christos Argyropoulos and Paloma A. Huidobro and Emanuele Galiffi and Fan Yang and John B. Pendry and David A. B. Miller},
  journal   = {Opt.\ Mater.\ Express},
  volume    = {15},
  number    = {7},
  pages     = {1544--1709},
  year      = {2025},
  publisher = {Optica Publishing Group},
  doi       = {10.1364/OME.559374}
}

@article{garciadeabajo2025roadmap,
  title     = {Roadmap for Photonics with 2D Materials},
  author    = {F. Javier {Garc\'{\i}a de Abajo} and D. N. Basov and Frank H. L. Koppens and Lorenzo Orsini and Matteo Ceccanti and Sebasti\'{a}n Castilla and Lorenzo Cavicchi and Marco Polini and P. A. D. Gon\c{c}alves and A. T. Costa and N. M. R. Peres and N. Asger Mortensen and Sathwik Bharadwaj and Zubin Jacob and P. J. Schuck and A. N. Pasupathy and Milan Delor and M. K. Liu and Aitor Mugarza and Pablo Merino and Marc G. Cuxart and Emigdio Ch\'{a}vez-Angel and Martin Svec and Luiz H. G. Tizei and Florian Dirnberger and Hui Deng and Christian Schneider and Vinod Menon and Thorsten Deilmann and Alexey Chernikov and Kristian S. Thygesen and Yohannes Abate and Mauricio Terrones and Vinod K. Sangwan and Mark C. Hersam and Leo Yu and Xueqi Chen and Tony F. Heinz and Puneet Murthy and Martin Kroner and Tomasz Smolenski and Deepankur Thureja and Thibault Chervy and Armando Genco and Chiara Trovatello and Giulio Cerullo and Stefano Dal Conte and Daniel Timmer and Antonietta De Sio and Christoph Lienau and Nianze Shang and Hao Hong and Kaihui Liu and Zhipei Sun and Lee A. Rozema and Philip Walther and Andrea Al\`{u} and Michele Cotrufo and Raquel Queiroz and X.-Y. Zhu and Joel D. Cox and Eduardo J. C. Dias and \'Alvaro {Rodr\'{\i}guez Echarri} and Fadil Iyikanat and Andrea Marini and Paul Herrmann and Nele Tornow and Sebastian Klimmer and Jan Wilhelm and Giancarlo Soavi and Zeyuan Sun and Shiwei Wu and Ying Xiong and Oles Matsyshyn and Roshan Krishna Kumar and Justin C. W. Song and Tomer Bucher and Alexey Gorlach and Shai Tsesses and Ido Kaminer and Julian Schwab and Florian Mangold and Harald Giessen and M. S\'anchez S\'anchez and D. K. Efetov and T. Low and G. G\'omez-Santos and T. Stauber and Gonzalo \'Alvarez-P\'erez and Jiahua Duan and Luis Mart\'{\i}n-Moreno and Alexander Paarmann and Joshua D. Caldwell and Alexey Y. Nikitin and Pablo Alonso-Gonz\'alez and Niclas S. Mueller and Valentyn Volkov and Deep Jariwala and Timur Shegai and Jorik van de Groep and Alexandra Boltasseva and Igor V. Bondarev and Vladimir M. Shalaev and Jeffrey Simon and Colton Fruhling and Guangzhen Shen and Dino Novko and Shijing Tan and Bing Wang and Hrvoje Petek and Vahagn Mkhitaryan and Renwen Yu and Alejandro Manjavacas and J. Enrique Ortega and Xu Cheng and Ruijuan Tian and Dong Mao and Dries Van Thourhout and Xuetao Gan and Qing Dai and Aaron Sternbach and You Zhou and Mohammad Hafezi and Dmitrii Litvinov and Magdalena Grzeszczyk and Kostya S. Novoselov and Maciej Koperski and Sotirios Papadopoulos and Lukas Novotny and Leonardo Viti and Miriam Serena Vitiello and Nathan D. Cottam and Benjamin T. Dewes and Oleg Makarovsky and Amalia Patan\`{e} and Yihao Song and Mingyang Cai and Jiazhen Chen and Doron Naveh and Houk Jang and Suji Park and Fengnian Xia and Philipp K. Jenke and Josip Bajo and Benjamin Braun and Kenneth S. Burch and Liuyan Zhao and Xiaodong Xu},
  journal   = {ACS Photonics},
  volume    = {12},
  number    = {8},
  pages     = {3961--4095},
  year      = {2025},
  publisher = {American Chemical Society},
  doi       = {10.1021/acsphotonics.5c00353},
}

@article{rodriguezecharri2026nonreciprocal,
  title     = {Nonreciprocal plasmons in one-dimensional carbon nanostructures},
  author    = {{\'A}lvaro {Rodr{\'\i}guez Echarri} and F. Javier {Garc\'{\i}a de Abajo} and Joel D. Cox},
  journal   = {Nat.\ Commun.},
  volume    = {17},
  number    = {1},  
  pages     = {1114},
  year      = {2026},
  publisher = {Nature Publishing Group UK London},
  doi       = {10.1038/s41467-025-67872-5}
}

@misc{SM,
  title = {See {S}upplemental {M}aterial at [link to be completed]},
}

@article{gueissaz1991high,
  title   = {High electron density and mobility in single and double planar doped {InGaAs}/{InAlAs} heterojunctions on {InP}},
  author  = {F. Gueissaz and R. Houdr{\'e} and M. Ilegems},
  journal = {J.\ Cryst.\ Growth},
  volume  = {111},
  number  = {1--4},
  pages   = {470--474},
  year    = {1991},
  doi     = {10.1016/0022-0248(91)91029-N}
}

@article{pozela2011electron,
  title   = {Electron transport in modulation-doped {InAlAs}/{InGaAs}/{InAlAs} and {AlGaAs}/{InGaAs}/{AlGaAs} heterostructures},
  author  = {J. Po{\v{z}}ela and K. Po{\v{z}}ela and V. Jucien{\.e} and A. Su{\v{z}}ied{\.e}lis and N. {\v{Z}}urauskien{\.e} and A. S. Shkolnik},
  journal = {Lith.\ J.\ Phys.},
  volume  = {51},
  number  = {4},
  pages   = {270--275},
  year    = {2011},
  doi     = {10.3952/lithjphys.51404}
}

@article{barker2005high,
  title   = {High-field electron transport in {AlGaN}/{GaN} heterostructures},
  author  = {J. M. Barker and D. K. Ferry and S. M. Goodnick and D. D. Koleske and A. Allerman and R. J. Shul},
  journal = {Phys.\ Status Solidi C},
  volume  = {2},
  number  = {7},
  pages   = {2564--2568},
  year    = {2005},
  doi     = {10.1002/pssc.200461384}
}
\end{document}